\documentclass[aps,prl,preprint,groupedaddress]{revtex4}
\usepackage{amsmath}
\usepackage{graphicx}
\usepackage{color}

\begin{document}

\title{Entanglement Control of Azobenzene by Photoisomerization
in NMR Quantum Computer}

\author{Taiga Yasuda$^1$, Masahito Tada-Umezaki$^2$, Mikio Nakahara$^{1,2}$ Tomonari Wakabayashi$^{2,3}$}

\affiliation{$^1$Department of Physics, Kinki University, 3-4-1 Kowakae, Higashi-Osaka, 577-8502, Japan \\
$^2$Research Center for Quantum Computing, Interdisciplinary Graduate School of Science and Engineering, Kinki University, 3-4-1 Kowakae, Higashi-Osaka 577-8502, Japan \\
$^3$Department of Chemistry, Kinki University, 3-4-1 Kowakae, Higashi-Osaka, 577-8502, Japan}

\begin{abstract}

Entanglement control of qubits in a photoisomerizing molecule is studied
in the context of an NMR quantum computer by taking azobenzene as 
an example.
Azobenzene has two different isomers,
{\it{}trans}-azobenzene (TAB) and {\it{}cis}-azobenzene (CAB), which
can be interconverted by photoisomerization.
Changing molecular structure leads to change in the spin-spin
coupling constant, and hence leads to change in entangling
operation time.
We first obtain stable structures of TAB and CAB by {\it ab initio}
calculation. 
Then, we calculate the NMR spectra of these isomers and verify
that they reproduce the chemical shift obtained experimentally with a
good precision.
Our result indicates that the coupling strength
between a $^{15}$N
and a $^{13}$C nuclei in the molecule changes by a large amount
under photoisomerization. 

\end{abstract}


\maketitle

\section{Introduction}

Quantum computer utilizes quantum mechanical phenomena, such as superposition and entanglement, as computational resources to outperform current digital 
computers. Nonetheless, a working quantum computer is yet to be realized. 
Feynman proposed the possibility of a quantum computer for the first
time \cite{feynman} and Deutsch triggered
quantum computer research by introducing a quantum Turing machine
\cite{deutsch}.
Shor's algorithm \cite{Shor} and Grover's algorithm \cite{Grover} 
demonstrated that a quantum computer can process some practical
problems more efficiently than a classical computer by making use
of an exponentially massive parallel processing, known  as
the quantum parallelism.

There are several physical realizations of a quantum computer proposed to date
\cite{roadmap,nakaharaohmi};
trapped ions \cite{ion2}, neutral atoms in optical lattices \cite{atom},
an NMR quantum computer
\cite{nmrreview1}, and superconducting qubits \cite{jj}, among others.
Nuclear spins in a molecule work as qubits, that
are controlled by nuclear magnetic resonance (NMR) spectrometer 
in an NMR quantum computer \cite{nmr1}.
An NMR quantum computer may have several qubits if an appropriate
molecules are employed. $^{13}$C-labelled chloroform \cite{chl1}
works as a two-qubit quantum computer, for example.

Controlling the coupling between nuclear qubits in a molecule by
structural change has been already attempted \cite{molswitch}.
In this paper, we analyze molecules whose inter-qubit coupling can be
controllable by photoisomerization.
Azo compounds are molecules whose two functional groups (aryl or alkyl)
R and R' are connected by an azo group (N=N) as R-N=N-R'. They
change their molecular structure 
between {\it{}trans}-azobenzene (TAB) and {\it{}cis}-azobenzene (CAB)
under photoisomerization. 
Application of the photoisomerization of azobenzene to
the switching device has been studied in \cite{iso1,iso2,iso3,iso4}.
In fact, CAB is unstable and it gradually changes to TAB, although 
the relaxation time of this process is long
enough to control qubits many times \cite{raman}.
One of the advantages of using azobenzene for quantum computation is that
it is possible to control interaction between qubits 
by changing molecular structure by UV light irradiation.
It is the purpose of this paper to study the interaction between nuclei 
in azobenzene and to propose the possibility of using azobenzene or similar
molecules as a multi-qubit NMR quantum computer with a tunable coupling.

\section{Theoretical Background}

Let us consider a molecule with two spin-1/2 nuclei in a strong magnetic field
{\boldmath$B_0$} along the $z$-axis.
We assume that the nuclear species of the spins are different.
The Hamiltonian of the molecule in the laboratory
frame is
\begin{equation}
\mathcal{H}= -\omega_{0,1} I_z \otimes I - \omega_{0,2} I \otimes I_z
+ \mathcal{H}_{\rm int},
\label{eq:labh}
\end{equation}
where $I_k = \sigma_k/2$ is the $k$th component of the Pauli spin matrix divided by 2 and $\omega_{0,i}$ is the Zeeman energy of the $i$th spin. 
The interaction Hamiltonian between the spins is given by
\begin{equation}
  \mathcal{H}_{\textrm{int}}=  \sum_{j,k
=x,y,z}J_{jk}{I}_{j}\otimes{} {I}_{k}.
\end{equation}
Here
$J_{jk}$ are called the spin-spin coupling constants (SSCC).
The coupling constants
 $J_{jk}$ take the isotropic Heisenberg form $J_{jk} = J \delta_{jk}$
for a molecule in a liquid state at room temperature, which we assume
to be the case throughout this paper. Then
$\mathcal{H}_{\textrm{int}}$ simplifies as
\begin{equation}
  \mathcal{H}_{\textrm{int}}=J\sum_{k=x,y,z}I_{k}\otimes{}I_{k}.
\end{equation}

Let us fix the basis vectors as
\begin{align}
  |0\rangle=
  \begin{pmatrix}
    1\\
    0
  \end{pmatrix},
\  |1\rangle=
  \begin{pmatrix}
    0\\
    1
  \end{pmatrix},
\end{align}
where $|0 \rangle$ ($|1 \rangle$) corresponds to the spin-up (spin-down)
eigenstate of $\sigma_z$. We take the order of the two-qubit binary
basis vectors
as $\{|00 \rangle, |01 \rangle, |10 \rangle, |11 \rangle\}$ as usual.
We transform the Hamiltonian in the laboratory frame (\ref{eq:labh})
to that in the rotating frame of each spin by introducing the unitary
transformation 
\begin{equation}
U (t) = e^{-i \omega_{0,1} I_z t} \otimes e^{-i \omega_{0,2} I_z t}.
\end{equation}
The Hamiltonian in the rotating frame is
\begin{eqnarray}
\mathcal{H}_{\rm rot} &= &U \mathcal{H} U^{\dagger} - i U \frac{d}{dt}
U^{\dagger}\nonumber\\
&=& J I_z \otimes I_z,
\end{eqnarray}
where rapidly oscillating matrix elements have been dropped.

The time-evolution operator in this frame is
\begin{eqnarray}
U_{\rm rot}(t) &=& e^{-i {\mathcal{H}}_{\rm rot} t}\nonumber\\
&=& \left( \begin{array}{cccc}
e^{-i J t/4}&0&0&0\\
0&e^{i J t/4}&0&0\\
0&0&e^{i J t/4}&0\\
0&0&0&e^{-i J t/4}
\end{array} \right),
\end{eqnarray}
where we have taken the natural unit, in which $\hbar = 1$.
Suppose the initial state is a product state $|+ 0 \rangle = \frac{1}{\sqrt{2}}(
|00 \rangle + |1 0 \rangle)$, for example, where $|\pm  \rangle$ are the
eigenvectors of $\sigma_x$ with the eigenvalues $\pm 1$.
This state may be generated by applying the Hadamard gate on
the first qubit of a state $|00 \rangle$, for example.
The action of $U_{\rm rot}(t)$ on $|+ 0 \rangle$ yields the state
\begin{eqnarray}
U_{\rm rot}(t) |+ 0 \rangle &=&\frac{1}{\sqrt{2}}\left(
e^{-i Jt/4}|00 \rangle + e^{i Jt/4}|10 \rangle\right)\nonumber\\
&=& \cos \left( \frac{Jt}{4} \right)|+0 \rangle
-i \sin \left( \frac{Jt}{4} \right)|-0 \rangle.
\end{eqnarray}
This shows that the spins are in a maximally entanged state (MES)
at $t=\tau$ such that
$|\cos (J\tau/4)|=|\sin (J\tau/4)|$, that is,
\begin{equation}
\tau =\frac{\pi}{|J|}(2n+1), \quad (n=0,1,2, \ldots).
\label{tau}
\end{equation}

\section{Method}


Stable structures and vibrational properties of azobenzene were already
reported in \cite{opt}, in which Gaussian03 with
density functional theory (DFT) and M{\o}ller-Plesset (MP) methods were employed. 
We used DFT method in the present work since MP method fails to calculate SSCC. 
First, we evaluated optimized structures of azobenzene with DFT calculation: 
B3LYP \cite{b3,lyp}, B3PW91 \cite{b3,pw2}, PW91PW91 \cite{pw2} and 
PBEPBE \cite{pbe2} with basis set: 6-31+G(d). 

Next, 
we calculated the NMR spectrum and SSCC of azobenzene using 
Gauge-Independent Atomic Orbital (GIAO) method \cite{giao1,giao4,spin1,spin2} 
and Integral-Equation-Formalism Polarizable Continuum Model (IEFPCM) method
\cite{pcm1,pcm4}. 
GIAO method was employed to calculate the NMR spectrum, while IEFPCM method was 
used to take account of the solvent effect of chloroform
in our calculation. 
Chemical shifts of carbons are relative to tetramethylsilane (TMS) 
and those of nitrogens are relative to ammonia (NH$_3$).

\section{Results}

\begin{figure}
  \includegraphics[width=0.5\linewidth]{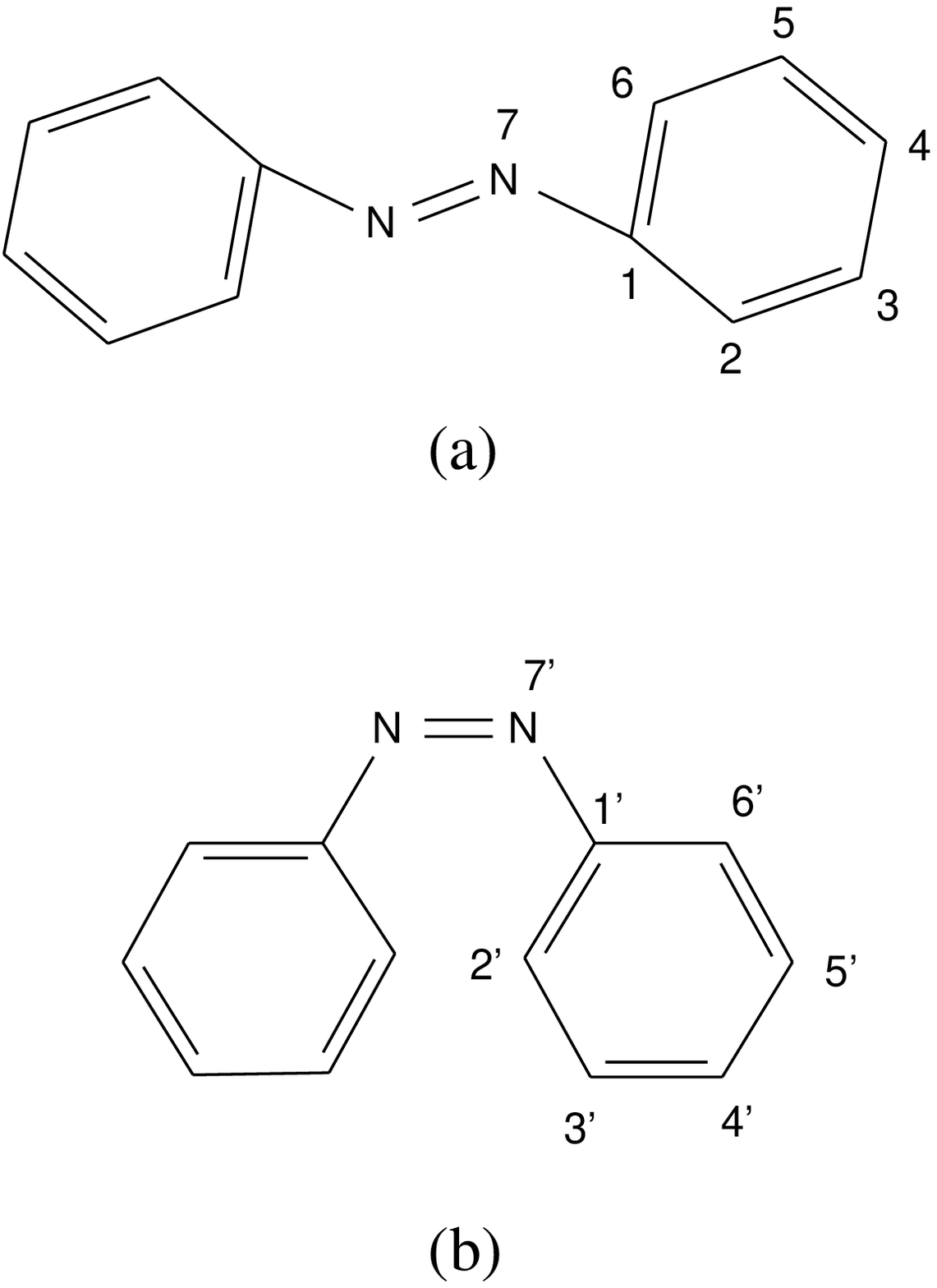}
  \caption{Schematic structures of {\it{}trans}-azobenzene (a) and {\it{}cis}-azobenzene (b).}
  \label{AB_eps}
\end{figure}

We calculated stable structures of azobenzene
with Gaussian03 with B3LYP/6-31+G(d), B3PW91/6-31+G(d), PW91PW91/6-31+G(d) and PBEPBE/6-31+G(d). 
The results obtained reproduce the previous ones \cite{opt} with a good
precision (data not shown). 
Figure~\ref{AB_eps} shows the schematic structures of TAB and CAB.
We studied entanglement between a nitrogen nucleus and a carbon
nucleus, which are denoted as nuclei 1 and 7 in Fig. \ref{AB_eps}~(a), and 
1' and 7' in Fig.~\ref{AB_eps}~(b), respectively. 
It should be noted that the N=N-C angle changes drastically by 
photoisomerization, under which the
molecule transforms between TAB and CAB.
We note from the NMR spectra of two nitrogen nuclei that they
are very weakly coupled,
that is $J \simeq 0$ for this pair, and hence it is takes extremely
long time to entangle these nuclei.

Table I shows the chemical shifts of the nitrogen 
and the carbon nuclei and SSCC between them we have obtained.
They are computed with DFT functionals with B3LYP/6-31+G(d), B3PW91/6-31+G(d), 
PW91PW91/6-31+G(d) and PBEPBE/6-31+G(d).
The times to attain the maximal entanglement were calculated with
Eq.~(\ref{tau}) by setting $n=0$.

Experimentally measured 
values of the chemical shifts of the carbon and the nitrogen
nuclei are also given in Table I \cite{spC}. 
The calculated chemical shifts of carbon nuclei
reproduce the experimental results
fairly accurately,
while those of nitrogen nuclei do not agree with the experimental results
with a good precision.
This difference might be attributed to
the difference in experimental conditions.
The observed chemical shifts of carbon nuclei were 
obtained for azobenzene in liquid chloroform solvent, 
while those of nitrogen nuclei were obtained for azobenzene in
a polycrystalline state \cite{spN1,spN2}.

Experimental values of SSCC between the nitrogen and the carbon
nuclei in azobenzene do not exist to our knowledge.
In contrast, the measured data of SSCC between carbon nuclei are available 
\cite{spN2} and they are in good agreement with our computational results
(data not shown).
We expect, from these evidences, that the computed SSCC between the nitrogen 
and the carbon nuclei are reliable. 
Time to produce MES in CAB is longer than that of TAB in all computational
results. The ratio of times required to attain MES in CAB
to that of TAB depends on the scheme; it is
4.2, 3.6, 2.3 and 2.4 with B3LYP, B3PW91, PW91PW91 and PBEPBE, respectively.

In this work, we have employed azobenzene to propose
the possible application of 
photoisomerizing molecules to an NMR quantum computer. 
The results showed that the time required to attain MES changes
by a large amount between the two isomers, CAB and TAB.
This suggests that photoisomerization can be used to control SSCC,
and hence entanglement in azobenzene, which is potentially
a powerful tool in NMR quantum computing.

\section{Conclusion and Discussion}

We calculated the time required to attain maximal entanglement
between the carbon and nitrogen nuclei in azobenzene with {\it ab initio}
methods.
As a result, the time to attain maximal entanglement in 
{\it cis}-azobenzene is approximately four times
longer than that in {\it trans}-azobenzene.
This reflects the fact that SSCC of
{\it cis}-azobenzene is four times weaker than that of {\it trans}-azobenzene.
We conclude that azobenzene can be employed as a molecule with
a tunable coupling in NMR quantum computing.

Search for photoisomerizing
molecules with a more drastic change in SSCC under 
photoisomerization is in progress and will be reported elsewhere.

\section{acknowledgement}

This work is supported by ``Open Research Center''
Project for Private Universities: Matching fund subsidy from
MEXT (Ministry of Education, Culture, Sports, Science and
Technology). MN's work is supported in part by Grant-in-Aid for Scientific
Research (C) from JSPS (Grant No. 19540422).

\clearpage

\clearpage

\begin{table*}
    \caption{Chemical shift, spin-spin coupling constant and time to attain 
maximally entangled state of {\it trans}-azobenzene (upper table) and
{\it cis}-azobenzene (lower table).
        $^{\rm a}$Chemical shift of nitrogen nucleus (7 and 7$'$) and
        $^{\rm b}$chemical shift of carbon nucleus (1 and 1$'$). 
Experimental values are from
      $^{\rm c}$Ref. \cite{spN1},
        $^{\rm d}$Ref. \cite{spC}, and
        $^{\rm e}$Ref. \cite{spN2}.
The value of $J(12,16)$ is an average of calculated $J(12)$ and $J(16)$
since the experimentally available value is an average of $J(12)$ and $J(16)$.
This also applies to $J(1'2',1'6')$.
}
    \begin{tabular}{llllll}
      \hline
       & N(7)$^{\rm a}$~[ppm] & C(1)$^{\rm b}$~[ppm] & $J(12,16)$~[Hz] & 
$J(17)$~[Hz] & $\tau$~[s] \\
      \hline
      B3LYP & 504 & 157 & 37 & -3.8 & 0.84 \\
      B3PW91 & 501 & 153 & 35 & -4.5 & 0.70 \\
      PW91PW91 & 486 & 157 & 33 & -8.9 & 0.35 \\
      PBEPBE & 486 & 156 & 33 & -8.5 & 0.37 \\
      Experiment & 509$^{\rm c}$ & 153$^{\rm d}$ & 34$^{\rm d}$ &  &  \\
    \end{tabular}

    \begin{tabular}{llllll}
      \hline
       & N(7$'$)$^{\rm a}$~[ppm] & C(1$'$)$^{\rm b}$~[ppm] & $J(1'2',1'6')$~[Hz]
& $J(1'7')$~[Hz] & $\tau$~[s] \\
      \hline
      B3LYP & 547 & 159 & 37 & -16 & 0.20 \\
      B3PW91 & 542 & 155 &36 & -16 & 0.20 \\
      PW91PW91 & 525 & 158 & 34 & -21 & 0.15 \\
      PBEPBE & 524 & 158 & 34 & -20 & 0.15 \\
      Experiment & 529$^{\rm e}$ & 154$^{\rm d}$ & 32$^{\rm d}$ &  &  \\
    \end{tabular}
    \label{nmr-jc}
\end{table*}


\begin{thebibliography}{99}
\bibitem{feynman}R. P. Feynman, ``Quantum Mechanical Computers'', Optics News (Feb., 1985) pp.11-20.
\bibitem{deutsch}D. Deutsch, Proc. R. Soc. London A {\bf{}400}, 97 (1985).
\bibitem{Shor}P. W. Shor, SIAM J. Comput. {\bf{}26}, 1484 (1997).
\bibitem{Grover}L. K. Grover, Phys. Rev. Lett. {\bf{}79}, 325 (1997).
\bibitem{roadmap} Quantum Information Science and Technology Roadmapping
project, http://qist.lanl.gov/
\bibitem{nakaharaohmi} M. Nakahara, and T. Ohmi,
{\it Quantum Computing: From Linear Algebra To Physical Realizations}, 
Taylor and Francis (2008).
\bibitem{ion2}M. \v{S}a\v{s}ura, and V. Bu\v{z}ek, J. Mod. Opt. {\bf{}49},
1593 (2002).
\bibitem{atom} I.~Bloch, Nature Physics, {\bf 1}, 23 (2005).


\bibitem{nmrreview1} L. M. K. Vandersypen and I. L. Chuang,
Rev. Mod. Phys. {\bf 76}, 1037 (2004).




\bibitem{jj} A.~Zagoskin and A.~Blais, Phys. in Canada {\bf 63}, 215 (2007).

\bibitem{nmr1}I. I. Rabi, J. R. Zacharias, S. Millman, and P. Kusch, Phys. Rev. {\bf{}53}, 318 (1938).

\bibitem{chl1}E. Knill, I. L. Chuang and R. Laflamme, Phys. Rev. A {\bf{}57}, 3348 (1998).
\bibitem{molswitch}G. A. Timco1, S. Carretta, F. Troiani, F. Tuna, R. J. Pritchard, C. A. Muryn, E. J. McInnes1, A. Ghirri, A. Candini, P. Santini, 
G. Amoretti, M. Affronte, and R. E. Winpenny, Nature Nanotechnology {\bf{}4}, 173 (2009).

\bibitem{iso1}C. M. Stuart, R. R. Frontiera, and R. A. Mathies, J. Phys. Chem. A {\bf{}111}, 12072 (2007).

\bibitem{iso2}I. Conti, M. Garavelli, and G. Orlandi, J. Am. Chem. Soc., {\bf{}130}, 5216 (2008).

\bibitem{iso3}F. Puntoriero, P. Ceroni, V. Balzani, G. Bergamini, and F. V$\ddot{o}$gtle, J. Am. Chem. Soc., {\bf{}129}, 10719 (2007).

\bibitem{iso4} T. Ikeda, and O. Tsutsumi, Science {\bf{}268}, 1873 (1995).


\bibitem{raman}C. M. Stuart, R. R. Frontiera, and R. A. Mathies, J. Phys. Chem. A {\bf{}111}, 12072 (2007).







\bibitem{opt}N. Kurita, S. Tanaka, and S. Itoh, J. Phys. Chem. A {\bf{}104 (34)}, 8114-8120 (2000).



\bibitem{lyp}C. Lee, W. Yang and G. Parr, Phys. Rev. B {\bf{}37}, 785-789 (1988).
\bibitem{b3}A. D. Becke, J. Chem. Phys. {\bf{}98}, 5648 (1993).

\bibitem{pw2}J. P. Perdew, J. A. Chevary, S. H. Vosko, K. A. Jackson, M. R. Pederson, D. J. Singh, and C. Fiolhais, Phys. Rev. B {\bf{}48}, 4979 (1993). 

\bibitem{pbe2}J. P. Perdew, K. Burke, and M. Ernzerhof, Phys. Rev. Lett. {\bf{}78}, 1396 (1997). 

\bibitem{giao1}R. McWeeny, Phys. Rev. {\bf{}126}, 1028 (1962). 

\bibitem{giao4}K. Wolinski, J. F. Hilton, and P. Pulay, J. Am. Chem. Soc. {\bf{}112}, 8251 (1990).

\bibitem{spin1}T. Helgaker, M. Watson, and N. C. Handy, J. Chem. Phys. {\bf{}113}, 9402 (2000).

\bibitem{spin2}V. Sychrovsky, J. Grafenstein, and D. Cremer, J. Chem. Phys. {\bf{}113}, 3530 (2000).

\bibitem{pcm1}M. T. Canc\`es, B. Mennucci, and J. Tomasi, J. Chem. Phys. {\bf{}107}, 3032 (1997).

\bibitem{pcm4}J. Tomasi, B. Mennucci, and E. Canc\`es, J. Mol. Struct. (Theochem) {\bf{}464}, 211 (1999).

\bibitem{spC}H. J. Shine, and W. Subotkowski, Mag. Res. Chem., {\bf{}29}, 964 (2005).
\bibitem{spN1}R. E. Wasylishen, W. P. Power, G. H. Penner, and R. D. Curtis, Can. J. Chem. {\bf{}67}, 1219 (1989).
\bibitem{spN2}R. D. Curtis, J. W. Hilborn, G. W. Michael, D. Lumsden, E. Wasylishen, and J. A. Pincock, J. Phys. Chem. {\bf{}97}, 1856 (1993).

\end{thebibliography}
\end{document}